\documentstyle[aps, eqsecnum, preprint]{revtex}
 \newcommand {\nc}{\newcommand}
 \nc{\eq}{\begin{equation}}
 \nc{\en}{\end{equation}}
 \nc{\eqa}{\begin{eqnarray}}
 \nc{\ena}{\end{eqnarray}}
 \nc {\norm}[1]{\parallel{#1}\parallel}
 \def\intg{{\cal Z}}

 \def\Drc{{\cal D}}
 \def\order{{\cal O}}
 \def\trans{{\cal T}}
 \def\unit{{\bf 1}}
 \def\ot{\otimes}
 
 \def\wtld{\widetilde}

 \def\etal{{\it et al} }
 \tighten
 \begin{document}
  \draft
  \parindent=0cm
  \title{A New Solution to Ginsparg-Wilson Relation from Generalized Staggered Fermion}
  \author{Jian Dai}
  \address{
   Room 2082, Building 48, Peking University, Beijing, P. R. China,
   100871\\
   daijianium@yeah.net}
  \author{Xing-Chang Song}
  \address{
   Theoretical Group,
   Department of Physics, Peking University,
   Beijing, P. R. China, 100871\\
   songxc@ibm320h.phy.pku.edu.cn}
  \date{February 12th, 2001}
  \maketitle
  \begin{abstract}
   A generalized anti-hermitian staggered Dirac operator is
   formulated. Its relation with noncommutative geometry is briefly
   reviewed. Once this anti-hermitian operator is modified to be
   ``$\gamma^5$-hermitian'', it will provide a new solution to
   Ginsparg-Wilson relation, basing on an abstract algebraic
   analysis of Neuberger's overlap construction and a redefinition of chirality.   \\

  \end{abstract}
  \section{Introduction}
   A long-lasting folklore that massless fermions are incompatible with lattice formalism of field theories\cite{NG} has
   been almost brought to an end due to the breakthroughs in 90's characterized in physical language by the introduction of infinitely many flavors \cite{neubergerREV}.
   Mathematically, the problem of chiral fermions on
   lattices has been addressed to {\it Ginsparg-Wilson relation}(GWR) \cite{GW}; and overlap
   solution devised by Neuberger \etal
   provides a promising candidate for the continuing numerical work \cite{overlap}. Intuitively speaking, overlap construction can be characterized by
   an operator function including a ``compactification'' of the argument operator. Dirac-Wilson operator with a heavy negative mass is chosen by Neuberger for the
   argument operator. In fact, we will present a generic
   algebraic analysis on the construction of this operator
   function in this letter, such that a large class of formal solutions to GWR
   is available. \\

   On the other hand, staggered fermion formalism has dominated the simulations of
   LQCD for a long time when chiral properties are concerned \cite{stagger0}; we found an intriguing relation between staggered Dirac
   operator and noncommutative geometry(NCG) recently \cite{dsStagger}. However, in our understanding, the traditional definition of chirality for staggered
   fermions which reserves a $U(1)$ chiral symmetry is unreasonable from a very simple physical argument. The main contribution of this letter is that
   a new solution to GWR can be found out, if a {\it generalized staggered Dirac operator}(GSDO, we will refer conventional
   staggered Dirac operator as SDO) with a heavy negative mass is taken as argument operator in the operator
   function mentioned above and a more reasoning alternative definition of chirality is
   adopted.\\

   This paper is organized as following. Anti-hermitian GSDO is defined in section
   \ref{secI}, then a brief review of the relation of this
   operator and NCG is given. Alternative
   definition of chirality can be fitted into GWR, unless the
   anti-hermitian GSDO is modified to be
   ``$\gamma^5$-hermitian''. In seciton \ref{secII}, an
   abstract algebraic analysis is forwarded to characterized the
   overlap solution to GWR, and GSDO can be matched into the solution class. Some open discussions
   are put in section \ref{secIII}.
  \section{Generalized Staggered Fermions, Noncommutative Geometry and ``Wrong'' Definition of
  Chirality}\label{secI}
   Define a GSDO on a 2m-dimensional lattice $\intg^{2m}$ as
   \eq\label{GSDO}
    \Drc_{GS}=\sum_{\mu=1}^{2m}(\gamma^\mu\ot\unit \partial_\mu +{ia\over 2}\gamma^{2m+1}\ot\gamma^\mu\Delta_\mu)
   \en
   in which we introduce notations as following: let $a$ be
   lattice constant, $T^\pm_\mu$ be shift operators in
   conventional meaning; $\partial_\mu={1\over 2a}(T^+_\mu-T^-_\mu)$,
   $\Delta_\mu={1\over a^2}(2-T^+_\mu-T^-_\mu)$;
   $\gamma^\mu$ consist of a set of hermitian gamma matrices in 2m-dimensional Euclidean space
   and $\gamma^{2m+1}=(-)^m\gamma^1\gamma^2\ldots\gamma^{2m}$. We hope to point out that $\{\gamma^\mu\ot\unit$,
   $\gamma^{2m+1}\ot\gamma^\mu: \mu=1,2,...,2m\}$ generate 4m-dimensional Clifford algebra
   $Cl(4m)$; in another word, let
   $\Gamma^\mu=\gamma^{2m+1}\ot\gamma^\mu$,
   $\Gamma^{2m+\mu}=\gamma^\mu\ot\unit$, and $\Gamma^\mu_\pm={1\over 2}(\Gamma^\mu\pm i\Gamma^{2m+\mu}),
   \mu=1,2,...,2m$,
   then GSDO can be put in a complex coordinate form
   \eq\label{sC}
    i\Drc_{GS}=\sum_{\mu=1}^{2m}(\Gamma^\mu_+ \partial^+_\mu +\Gamma^\mu_- \partial^-_\mu)
   \en
   where $\partial^\pm_\mu=\frac{1}{a}(T^\pm_\mu -1)$.
   It is easy to verify that $\Drc_{GS}$ satisfies the ``square-root'' condition
   $\Drc_{GS}^2=\Delta$ where lattice laplacian $\Delta=-\partial^+_\mu\partial^-_\mu=-\sum_\mu\Delta_\mu$,
   and that $\Drc_{GS}^\dag =-\Drc_{GS}$.   \\

   It has been shown in \cite{stagger} that definition (\ref{GSDO}) is equivalent to
   SDO adopting {\it staggered phases}
   $\eta_\mu(x)$ as gamma matrices when 2m=4 and Clifford algebra $Cl(8)$ stands in a specific
   representation. We have given a more explicit proof to this equivalency in cases 2m=2,4 in \cite{dsStagger}.
   Here two points shall be clarified.
   The first one is that when 2m=2,4, the difference between staggered phase formalism and GSDO
   is that no peculiar representation is specified in definition (\ref{GSDO}) or (\ref{sC}), so that generically GSDO may
   have no ``stagger'' interpretation as a geometric picture.
   The second one is that the equivalence of SDO and GSDO in
   higher even-dimensional cases has not yet been proved.\\

   A deep relation between GSDO and NCG has been discovered in our recent
   work and we give a outline of this relation in this paragraph.
   Slightly modify Eq.(\ref{sC}) to make the following definition
   of a ``geometric Dirac operator''
   \[
    F:=\sum_{\mu=1}^{2m}(\Gamma^\mu_+ T^+_\mu +\Gamma^\mu_- T^-_\mu)
   \]
   $F$ possess follow properties:\\
   i) Connes' distance formula $d_F(x,y)=\sup \{|f(x)-f(y)|:\norm{[F,f]}\leq 1\}$, $\forall x,y\in
   \intg^{2m}$\cite{DISformula}
   endows an induced metric on lattices. We proved that when 2m=2,
   $d_F(x,y)=\sum_{i=1}^2(x^i-y^i)^2, \forall x,y\in \intg^2$ in \cite{dsDIS}, which recovers Euclidean geometry
   on 2D lattices. Note that since the difference part of $F$ and $\Drc_{GS}$ commutes with any $f$,
   $d_F(,)=d_{\Drc_{GS}}(,)$. \\

   ii) $i[F,f]$ provides an involutive representation of differential form $df$ on
   lattices under Connes' NCG construction. This representation
   can be extended to higher forms and due to that $F$ is an {\it index-zero Fredholm operator}
   subjected to the so-called ``geometric square-root'' condition $F^2\sim 1$,
   this extension is ``Junk-free'' \cite{dsDC}.\\

   iii) Gauge coupling is added by twisting $F$ with link variables $\omega$
   \[
    F(\omega):=\sum_{\mu=1}^{2m}(\Gamma^\mu_+ \omega_\mu^\dag T^+_\mu +\Gamma^\mu_- T^-_\mu \omega_\mu)
   \]
   Then Wilson action for lattice gauge fields can be expressed as
   \eq\label{act}
    S[\omega]=Tr((F(\omega)\wedge F(\omega))(F(\omega)\wedge F(\omega)))
   \en
   with a properly-defined wedge product, providing that $\omega$ is unitary \cite{dsWilson}.\\

   Let $\epsilon=\gamma^{2m+1}\ot\gamma^{2m+1}$, then it is obvious that
   $\{\Drc_{GS},\epsilon\}=0$. Till now chirality for SDO is defined to be $\epsilon$;
   accordingly, staggered formalism has a $U(1)_A$ chiral symmetry which is the main advance of this
   theory. However, we understand that $\gamma:=\gamma^{2m+1}\ot\unit$ is a more meaningful
   choice for chirality in physics. In fact, consider continuum limit(C.L.) in Eq.(\ref{GSDO}),
   $\Drc_{GS}\stackrel{C.L.}{\rightarrow}\gamma^\mu\ot\unit\frac{\partial}{\partial
   x^\mu}$, and $\gamma$ is nothing but the chirality in continuum
   theory. Of course on a lattice, $\{\Drc_{GS},\gamma\}\neq 0$. But
   it has been a general belief that GWR provides a final answer to the problem of chiral
   fermions on lattices. Hence, what we plan to do next is to
   define a lattice Dirac operator $\wtld{\Drc}_{GS}$ such that
   GWR holds for $\wtld{\Drc}_{GS}$ and $\gamma$. As the first
   step, the anti-hermitian $\Drc_{GS}$ must be modified to be a
   $\gamma$-hermitian $D_{GS}$ defined as
   \eq\label{gGSDO}
    D_{GS}=\sum_{\mu=1}^{2m}(\gamma^\mu\ot\unit \partial_\mu +{a\over 2}\gamma^{2m+1}\ot\gamma^\mu\Delta_\mu)
   \en
   such that $\gamma D_{GS}\gamma=D_{GS}^\dag$.
  \section{Solution to Ginsparg-Wilson Relation from generalized staggered Dirac
  Operator}\label{secII}
   The most popular solution to GWR now is overlap operator devised by
   Neuberger. We will describe Neuberber's construction in a pure algebraic
   manner so that generalizations are easy to be searched out. Note that as in most physical literature,
   mathematical rigidity on convergency is not cared in detail. \\

   GWR can be abstracted as
   a operator pair $(D,\Gamma)$ and a constraint relation
   \eq\label{aGWR}
    \{D, \Gamma\}=aD\Gamma D
   \en
   where $\Gamma$ satisfies $\Gamma^2=1,\Gamma^\dag=\Gamma$ and $D$ is $\Gamma$-hermitian $\Gamma D\Gamma=D^\dag$.
   Any solution can be written in the form $aD=1+V$; thus it is
   easy to verify that Eq.(\ref{aGWR}), together with the properties of $D,\Gamma$, imply $V$ is $\Gamma$-hermitian and
   unitary. On the other hand, any $\Gamma$-hermitian unitary
   operator $V$ will provide a formal solution to
   Eq.(\ref{aGWR}).\\

   Define a compactification transformation $\trans$ of operators to be
   $\trans(A)=A(A^\dag A)^{-\frac{1}{2}}$ for a generic operator
   $A$. We claim these three statement hold:
   \eqa
   \label{T1}
    \hbox{$\trans(A)$ is unitary;}\\
   \label{T2}
    \trans(A)=(AA^\dag)^{-\frac{1}{2}}A;\\
   \label{T3}
    \Gamma A\Gamma=A^\dag\Rightarrow
    \Gamma\trans(A)\Gamma=\trans(A)^\dag
   \ena
   In fact, from knowledge of linear algebra, $A$ can be
   decomposed as $A=U\lambda V^\dag$ where $U,V$ are two unitary
   operators and $\lambda$ is a diagonal operator, so $A^\dag=V\bar{\lambda} U^\dag$. Based on
   these facts, any $\alpha$-order power of $A^\dag A$ and $AA^\dag$ are
   defined to be $(A^\dag A)^\alpha=V|\lambda|^{2\alpha}V^\dag$,
   $(AA^\dag)^\alpha=U|\lambda|^{2\lambda}U^\dag$.
   Hence, both $(A^\dag A)^\alpha$ and $(AA^\dag)^\alpha$ are
   hermitian. Then $\trans(A)\trans(A)^\dag=A(A^\dag A)^{-\frac{1}{2}}(A^\dag
   A)^{-\frac{1}{2}}A^\dag=A(A^\dag A)^{-1}A^\dag=1$, and $\trans(A)^\dag \trans(A)=(A^\dag A)^{-\frac{1}{2}}A^\dag A(A^\dag
   A)^{-\frac{1}{2}}=1$, from which Eq.(\ref{T1}) follows.
   $(AA^\dag)^{-\frac{1}{2}}A=U\lambda|\lambda|^{-1} V^\dag=A(A^\dag
   A)^{-\frac{1}{2}}$, i.e. Eq.(\ref{T2}). Eq.(\ref{T3}) can be shown by $(\Gamma
   A\Gamma=A^\dag)\Rightarrow (\Gamma A^\dag
   A\Gamma=AA^\dag)\Rightarrow([U^\dag\Gamma V, |\lambda|^2]=0)\Rightarrow ([U^\dag\Gamma V, |\lambda|^{2\alpha}]=0)\Rightarrow
   (\Gamma (A^\dag
   A)^\alpha\Gamma=(AA^\dag)^\alpha)\stackrel{(\ref{T2})}{\Rightarrow}
   \Gamma\trans(A)\Gamma=\trans(A)^\dag$. Consequently, an
   operator function
   $\textbf{D}(A):=(1+\trans(A))/a$ provides a formal solution to Eq.(\ref{aGWR}), if $A$ is
   $\Gamma$-hermitian.\\

   $A$ is chosen by Neuberger to be Dirac-Wilson operator $\Drc_W$
   with a negative mass $-1/a$, and $\Gamma=\gamma^5$. We can
   continue to abstract and to generalize this construction to a finite summation decomposition $A=-\frac{1}{a}+A_0+aA_1+a^2A_2+\ldots +a^NA_N$ for some natural number $N$
   in which $A_i,i=1,2,\ldots, N$ are
   $\order(1)$ operators under C.L. and $A_{even}^\dag=-A_{even},\Gamma
   A_{even}\Gamma=-A_{even}$, $A_{odd}^\dag=A_{odd},\Gamma A_{odd}\Gamma=A_{odd}$. Hence the $\Gamma$-hermiticity of $A$ follows, and
   $\textbf{D}(A)\stackrel{C.L.}{\rightarrow} A_0$.
   As a physical requirement, $A_0$
   has to approach $i\gamma^\mu p_\mu$ in momentum representation under C.L.. More subtleties on the choices of other $A_i$ have to be taken
   care to make $\textbf{D}(A)$ meaningful in physics. In the case of
   $A=\Drc_W$, $A_0=\gamma^\mu\partial_\mu$, $A_1=\Delta/2$. \\

   Now let $A=D_{GS}-\frac{1}{a}$, and above abstract algebraic construction can be specified as
   $\Gamma=\gamma^{2m+1}\ot\unit=\gamma$, $N=1$,
   $A_0=\gamma^\mu\ot\unit \partial_\mu$, $A_1={1\over 2}\gamma^{2m+1}\ot\gamma^\mu\Delta_\mu$. Then a solution to GWR for a 2m-dimensional lattice
   is
   \[
    \wtld{\Drc}_{GS}=\textbf{D}(D_{GS})=\frac{1}{a}(1+\trans(D_{GS}))
   \]
   It is easy to check that $\wtld{\Drc}_{GS}$ has continuum limit
   $i\gamma^\mu\ot\unit \frac{\partial}{\partial x^\mu}$.
  \section{Discussions}\label{secIII}
   i) The old problem concerning staggered formalism of the $2^m$ flavor
   doubling is still presented in our formulation as being shown
   in the C.L. of $\wtld{\Drc}_{GS}$.\\

   ii) An physically-admissible massless Dirac operator $\Drc$ on a 4-dimensional lattice shall
   fulfill the following conditions:
   \begin{enumerate}
    \item ({\it Locality}) in moment space, $\wtld{\Drc}(p)$ is an analytic function of
    momentum $p$;
    \item ({\it No doubling}) $\wtld{\Drc}(p)$ is invertible for any $p$, except $p=0$;
    \item ({\it Continuum limit}) $\wtld{\Drc}(p)\sim i\gamma^\mu p_\mu
    +\order(ap^2)$, when $p\rightarrow 0$;
    \item ({\it Chirality}) GWR holds for $\Drc$
    \[
     \{\Drc, \gamma^5\}=a\Drc\gamma^5\Drc
    \]
   \end{enumerate}
   Only condition 3,4 are considered in this letter, and it is in
   proceeding whether condition 1,2 are satisfied by
   $\wtld{\Drc}_{GS}$.\\

   iii) It can be referred as a ``Wick rotation'' for the
   modification from $\Drc_{GS}$ to $D_{GS}$ and it is a profound
   change since the relations of the former to NCG that we outlined
   in Sec.\ref{secI} can not be expected to be hold for the
   latter. However, LQCD action can be put in this form
   \[
    S[\omega, \psi, \overline{\psi}]=Tr((F(\omega)\wedge F(\omega))(F(\omega)\wedge
    F(\omega)))+(\psi, \wtld{\Drc}_{GS}\psi)
   \]
   compared with Eq.(\ref{act}).\\

  {\bf Acknowledgements}\\
    This work was supported by Climb-Up (Pan Deng) Project of
    Department of Science and Technology in China, Chinese
    National Science Foundation and Doctoral Programme Foundation
    of Institution of Higher Education in China. J.D. expects to
    express his gratitude to Professor Chuan Liu for the helpful
    and inspiring discussion on GWR.

 \end{document}